\documentclass[11pt]{article}
\usepackage{graphicx}
\usepackage{csquotes}

\begin{document}

\title{On machine learning search for gravitational lenses}
\author{H.G. Khachatryan$^{1,2}$ \\
$^{1}$\footnotesize{Center for Cosmology and Astrophysics, Alikhanian National Laboratory}\\ 
$^{2}$\footnotesize{Yerevan State University, Yerevan, Armenia}
}

\maketitle

\begin{abstract}
We consider a machine learning algorithm to detect and identify strong gravitational lenses on sky images. First, we simulate different artificial but very close 
to reality images of galaxies, stars and strong lenses, using six different methods, i.e. two for each class. Then we deploy a convolutional neural network 
architecture to classify these simulated images. We show that after neural network training process one achieves about 93 percent accuracy. As a simple test for 
the efficiency of the convolutional neural network, we apply it on an real Einstein cross image. Deployed neural network classifies it as gravitational lens, thus 
opening a way for variety of lens search applications of the deployed machine learning scheme.
\end{abstract}

\section{Introduction}
One of the striking outcomes of General Theory of Relativity is that any gravitating body attracts light in specific way, the effect known as gravitational lensing. 
Commonly, the gravitational lensing effect is classified as strong, weak and micro. Strong gravitational lensing is the effect able to produce multiple images, such 
as Einstein cross, arcs or rings. It is believed that almost any light from distant galaxies reach us affected by any class of gravitational lensing effect. Although 
it was emphasized already by Zwicky \cite{zwicky1937} that "the probability that nebulae which act as gravitational lenses will be found becomes practically certainty", 
there were no direct observations of lensed objects until the second half of last century. The first multiple imaged object observed and claimed as a strong 
gravitational lens was QSO 0957+561  \cite{walsh1979}.

Among modern computer science's rapidly evolving technologies the machine learning (ML) is efficiently used in many areas of science. ML provides computational algorithms 
that improve the accuracy through experience \cite{mitchell1997} and is already used in astrophysical problems e.g. for CMB foreground reducing \cite{petroff2020}, search 
of AGN and pulsars in Fermi catalog \cite{zhu2020} and other cases \cite{diego2020,drozdova2020,isanto2018}. 

Regarding the gravitational lensing, the most attractive application for ML tools is the possibility for automatic detection of strong gravitational lenses (GL)
\cite{delchambre2019,huang2020,metcalf2019,mirzoyan2019,paraficz2016,petrillo2019,schaefer2018}. One of the main difficulties of this kind of application is that there 
is not enough big sample available by now (about 200 objects) of real observed strong GL. While to use the popular convolutional neural network (CNN), traditionally 
used for image recognition, one needs about thousands or more GL samples. So, one of the possible ways to deal with this problem is to generate artificial images of 
lenses, having, however, the known real lens image parameters, like sky level, telescope noise and source to lens layout in the space \cite{kitching2012} and then apply 
that model to real images. Here we combine the methods of simulation of strong GLs with modern ML methods to detect them. As an outcome we have fully functional ML model 
that can be applied for real all sky surveys like SDSS, HST, KIDS, CFHTLS and others \cite{ahumada2019,kuijken2019,ratnatunga1999}.

The paper is organized as follows. At first we simulate artificial images, then construct a CNN to detect strong GL among them. The main results are discussed in the last 
section.

\section{Data simulation}

For data simulation we use two python packages \textit{pyautolens} and \textit{lenstronomy} \cite{birrer2018, nightingale2018, nightingale2019}. \textit{pyautolens} package 
was created to model and estimate real strong gravitational lens parameters, but it also has capabilities to simulate artificial images of sky regions with galaxies, stars 
and strong gravitational lenses by employing its unique ray tracing algorithm. Because \textit{pyautolens} was able to reconstruct mass distribution around real strong 
gravitational lenses \cite{nightingale2018}, then it is first choice tool to simulate strong GL-s, too. Both \textit{pyautolens} and \textit{lenstronomy} reconstruct strong 
GL-s in as many details as possible. They reconstruct even telescope side effects like convolution, noise and pixelization. For ML methods it is important that negative cases 
(in our case stars and galaxies) also have similar digital properties like positive ones (i.e. noise, pixelization). Therefore we should use that two packages for simulating 
stars and galaxies, too. For every of these three type of astrophysical objects (stars, lenses, galaxies) we create 2 methods to simulate artificial images of them and these 
six methods allows one to have very random sample ready to be used in any ML algorithm. So we use it to simulate 2500 set of random stars, strong gravitational lenses, elliptic 
and spiral galaxies. We used well-known empiric Sersic profile
\begin{equation}
\mu(r)=\mu_{0}e^{-\beta(\frac{r}{r_e})^{1/n}},
\label{sersic}
\end{equation}
($r_e$ is a scaling radius, $\beta$, n are constants) for surface intensity of galaxies to simulate elliptic and spiral galaxies along with \textit{pyautolens} package.  We are 
going to construct very simple ML model with parameters number of free parameters $n\approx10000$, therefore having dataset of size 15000 examples more than enough to 
accomplish the task.

\begin{figure}[htbp]
	\centering
		\includegraphics[width=0.8\textwidth]{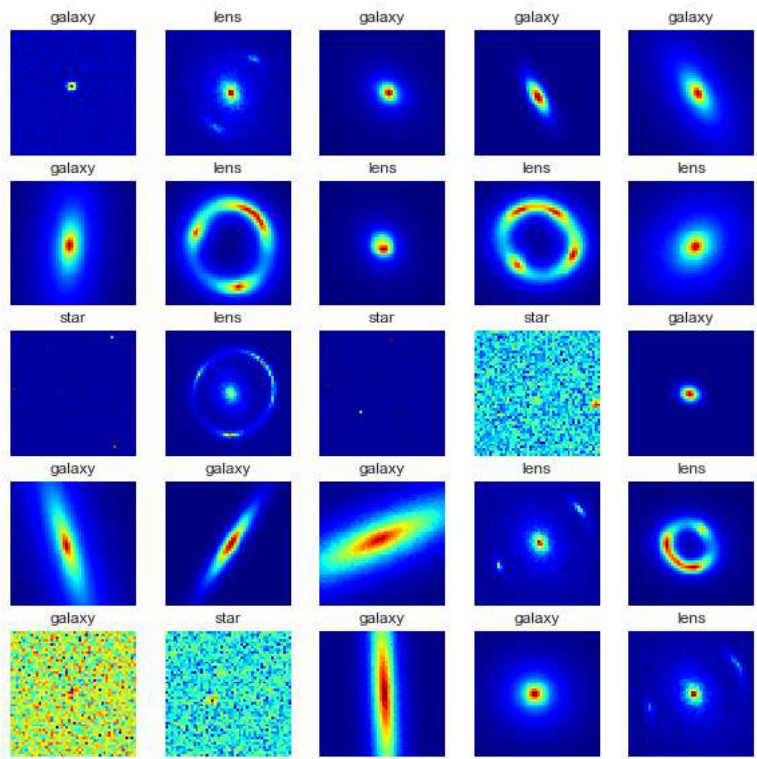}
	\caption{A sample of 25 simulated images of stars, galaxies and lenses.}
	\label{sim}
\end{figure}

Although \textit{pyautolens} gives realistic sky region images of strong gravitational lenses, it has a crucial problem, as it does not enable to generate the diversity 
of images of gravitational lens, of galaxy or of cluster of galaxies. In that case, another python package called \textit{lenstronomy} is used to generate 2500 set of 
sky regions with images of strong gravitational lenses. For having a truly random component in our data, we generated 2500 regions of sky with stars with by simply 
adding random five bright points on Gaussian random field. A sample of that simulation one can see on Fig.(\ref{sim}). 

There is so-called Bologna Lens Factory (BFS) Challenge \cite{schaefer2018} that provides a very large amount of simulated lens data, but the aim of that challenge is 
to discover gravitational lenses of different kind but here we focused on strong gravitational lenses. The BFS challenge is much more hard to accomplish since sometimes
it is even hard for human to distinguish between ordinary galaxy and lensed star.

\section{Neural network design}

Artificial neuron network or simply neural network (NN) is a mathematical model that mimics the properties of a biological neuron. McCulloch and Pitts \cite{mccullock1943} 
proposed a simple mathematical model for a neuron with inputs $x_j$ and output $y$
\begin{equation}
y=\theta\left(\sum_{j=1}^{n}{w_j}{x_j}-{w_0}\right),
\label{nn}
\end{equation}
here $\theta(x)$ is so-called activation function, and $w_j,\,w_0$ are the weights and the bias, respectively. In their work McCulloch and Pitts \cite{mccullock1943} used a 
threshold function as activation function, which is equivalent to Heaviside step function. In that case, the neuron did not transmit any signal till the combination under the 
sum operator had been lower than given fixed value. 

To construct a NN one can place these mathematical neurons on vertices of a graph with given topology, so that the edges of that directed graph represent the connection between 
neurons. Commonly, neurons are grouped in so-called layers where every neuron has no connection to another neuron in the same layer. Also, every NN should have so-called input 
and output layers by which it interacts with environment and the signal is transmitted only from the input layer to the output layer. Layers that are placed between the input 
and output layers are commonly called inner layers. For having a functional NN it should have at least one inner layer. Any NN that has more than one inner layer is called deep 
neural network.

The solid mathematical basis for NNs was established by Kolmogorov and Arnold \cite{kolm1956,kolm1957,arnold1957}. In that series of works they proved so-called universal 
approximation theorem. It states that any multivariate continuous function $f(x_{1},\dots,\,x_{n})$ can be represented by a finite composition of continuous single variable 
functions
\begin{equation}
f(x_{1},\dots,\,x_{n})=\sum_{q=0}^{2n}\Phi_{q}\left(\sum_{p=1}^{n}\psi_{p,q}(x_p)\right).
\label{kolm}
\end{equation}
This theorem \cite{kolm1957} ensures that any multivariate function can be approximated by some NN. Numerical simulations showed that a linear function can be approximated with 
NN that has only one inner layer and that the deep neural network can approximate a non-linear one. So for complex problems, it is preferable to use deep neural networks.

Then the paper of Hubel and Wiesel \cite{hubel1959} (marked by Nobel prize in physiology) on the structure and function of visual cortex was instrumental in inventing a new type 
of layers in NNs that can efficiently recognize objects on images. That new type of neurons called convolutional. That NN that has at least one convolutional layer is called 
convolutional neural network (CNN). Other types of layers that we use here, are max-pooling and dense. Max-pooling layer simply propagates maximal value of sub-array of a given size. 
The dense layer is an ordinary layer consisting of neurons proposed by \cite{mccullock1943}. We use flatten operation (sometimes called layer, too) for just transforming any matrix or
tensor into a vector.

\begin{figure}[htbp]
	\centering
		\includegraphics[width=0.8\textwidth]{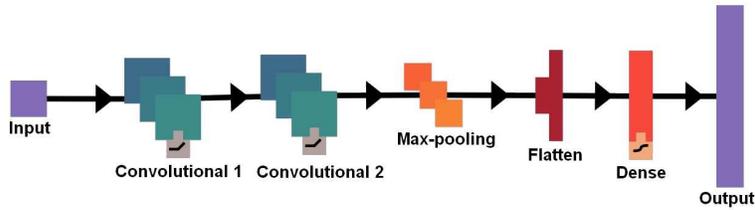}
	\caption{Convolutional neural network architecture.}
	\label{CNN}
\end{figure}

Here we use 7-layer deep CNN with leaky rectified linear unit function (RELU) as activator. The first two inner layers are convolutional ones with 32 and 16 neurons, respectively. 
Then comes the max-pooling layer with pool size 2x2. For the output layer we have simple dense layer with 3 neurons equal to the number of image classes. Architecture of the CNN 
is depicted on Fig.(\ref{CNN}). 
\begin{figure}[!htbp]
	\centering
		\includegraphics[width=0.8\textwidth]{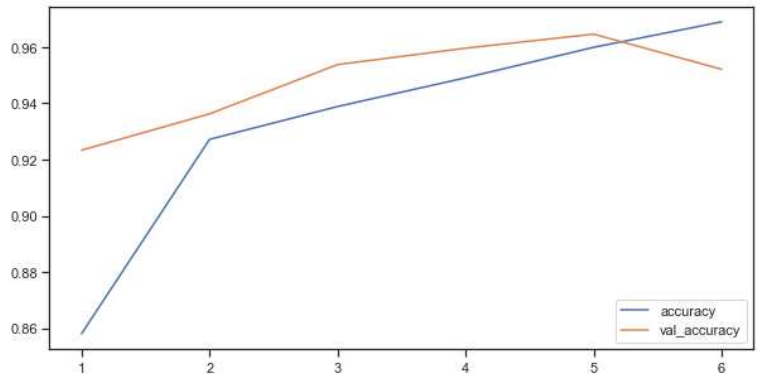}
	\label{learning}
	\caption{Neural network learning curve: dependence on the epoch for accuracy of both validation and train sets. One can see that accuracy for training set increases with every 
	         epoch run but after 5th epoch neural network goes into stage of over-fitting when accuracy for validation set drops, so the learning process stops automatically.}
\end{figure}

\begin{figure}[htbp]
	\centering
		\includegraphics[width=0.8\textwidth]{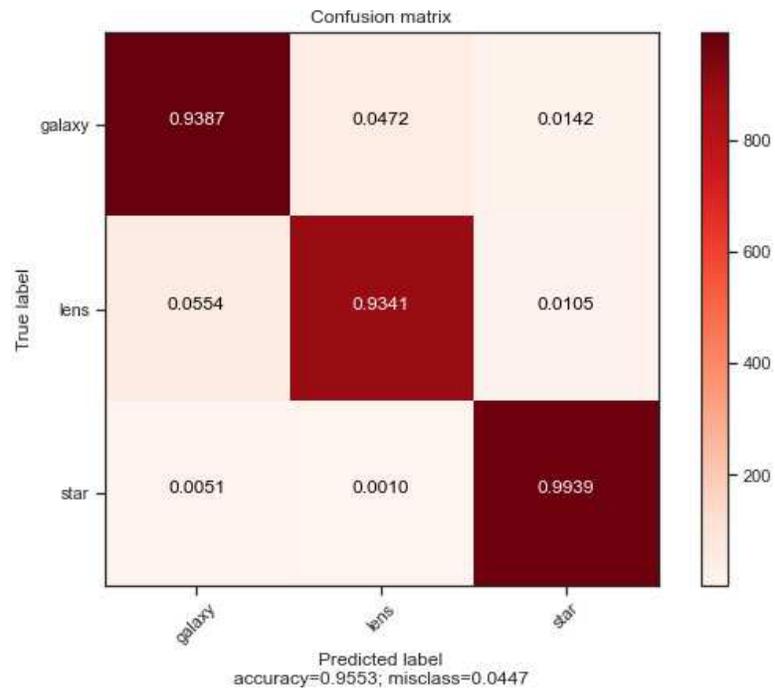}
	\caption{The confusion matrix for the test set. It can be seen that, the convolutional neural network works quite satisfactory for all categories. The worst accuracy
	         is achieved for the lenses (i.e $93.51\%$).}
	\label{confmat}
\end{figure}

For machine learning models a very simple measure for accuracy is defined. It is the ratio of the number of correctly predicted labels to the true labels. To see the accuracy of 
the CNN model, one needs to construct the confusion matrix (see Fig.(\ref{confmat})). The rows of the confusion matrix represent true labels of data, i.e. in our case star, galaxy 
and lens. And columns represent labels predicted by machine learning model. So, the diagonal elements represent cases when predicted and true labels coincide. One can calculate 
accuracy for every class and see that the worse accuracy we get in this model is for the lens class $93.51\%$.

Any ML model parameters are fixed through so called training process. In case of CNN model during training process an image is provided for input layer and for output it is expected
that the model should be able to reveal corresponding image type: i.e. star, lens or galaxy in our case. For CNN training of model is done by a technique called back-propagation 
\cite{mitchell1997}. In ML epoch is called a period of training process when all input data passes through a model. Practice shows that for reliable training more than one training 
epoch is needed. But too many training epoch can cause a situation when model just memorize input data properties and could be useless for any new unseen data. That kind of situation
in ML model called over-fitting. To avoid over-fitting, we split randomly whole dataset to two sets. We take 90 percent of examples of original dataset as training set and 10 
percent as validation set. We do training process on training set and check results on validation set. There are many regularization methods that helps to avoid over-fitting. One of 
them is so called early stopping. It monitors development of some properties of training and validation sets during training process and decides to terminate it by some predefined rules. 
Here we use very strict and simple early stopping rule. We stop training process as soon as model accuracy on training set becomes higher than on validation one. On Fig.(\ref{learning}) 
we depicted graph for model accuracy versus epoch number. As one can see after some epochs validation set accuracy decreases and that can be sign of over-fitting so the training process 
terminated after that epoch. 

Training process nowadays is done by gradient descent method to find a minimum for cost function. Gradient descent method introduces so called learning rate parameter, that show how 
fast the model parameters will be changed depending on gradient value. Then another problem of any ML model arises so called learning rate parameter tuning. When one set it too small it 
becomes possible to wander around local minimum. On other side, when it is set too large the it is possible to skip general minimum, too \cite{mitchell1997}. But nowadays there are too 
many methods of adaptive learning rate, that helps to mitigate that problem. Here we use adaptive momentum estimation (ADAM) method to solve that problem.

\section{Lens detection}

\begin{figure}[htbp]
	\centering
		\includegraphics[width=0.8\textwidth]{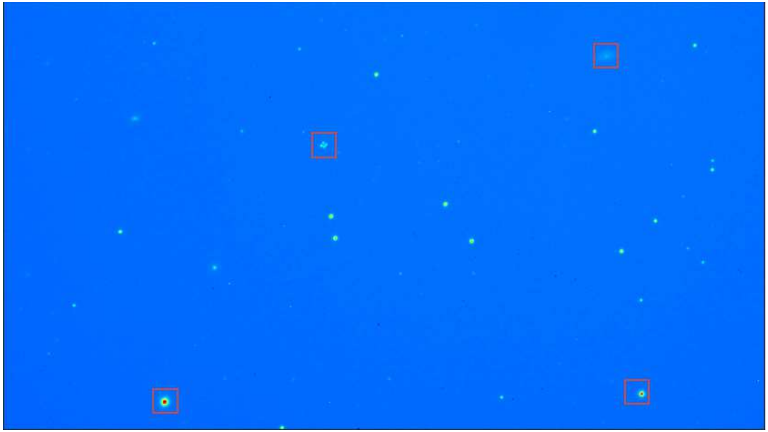}
	\caption{The image of HE 0435-1223 strong gravitational lens acquired by FORS2 instrument of VLT. Red boxes indicate selected regions for which CNN is applied to 
	         get predictions: galaxy, lens or star. See the next image for more details.}
	\label{eso}
\end{figure}

To test the neural network functionality on real images we have used the image acquired by Very Large Telescope (VLT) of well-known strong gravitational lens HE 0435-1223 \cite{wisotzki2002}. 
The image is shown on Fig.(\ref{eso}) was observed by FORS2 instrument of VLT. Besides the prominent strong gravitational lens one can see also some stars, galaxies and other 
objects on the image. We take four regions with some bright objects among them a region containing the gravitational lens. CNN was trained on images of size 50x50 pixels, so that regions 
should have the same size, otherwise no prediction can be provided.

\begin{figure}[htbp]
	\centering
		\includegraphics[width=0.8\textwidth]{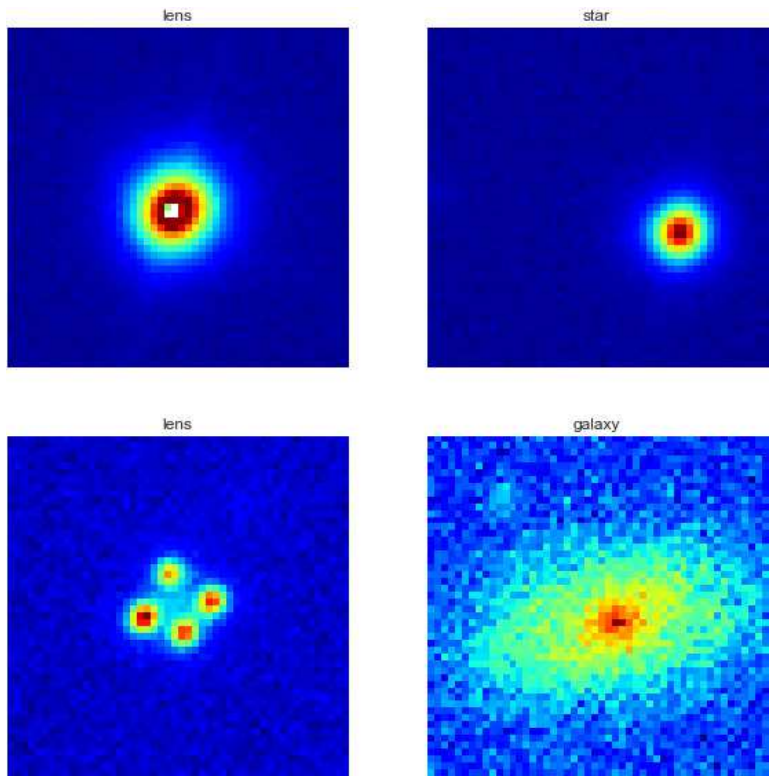}
	\caption{Predicted labels for four selected regions. One can see that, the CNN predicted labels not always match the reality, but the strong gravitational lens, 
	         i.e. Einstein cross, it predicted, correctly.}
	\label{regs}
\end{figure}

As one can see for selected random 4 regions, the CNN predicted one as a galaxy, correctly (see Fig.(\ref{regs})). Two regions are predicted as lenses, but only one of those two regions is really a 
gravitational lens. And finally one region is predicted as star. Although, one region was incorrectly classified as a lens and it can mean bad accuracy ($\approx50\%$) for real 
GLs but this deployed CNN can be used as a first step in very tedious process of searching real lenses on a big area of the sky. 

\section{Conclusions}

We have simulated images of artificial galaxies, stars and gravitational lenses and for them we constructed a convolutional neural network, aimed to identify those types of images.
We showed that even with very simple CNN model one can easily identify each type of image with an accuracy ($\approx93\%$). This opens a way regarding larger samples of gravitational 
images, to construct such CNN and detect gravitational lenses. 

Namely, we used CNN to predict the object types for real observations. As shown, although the artificial data are used to train CNN, it enables to predict correctly the object types,
especially for real gravitational lens. As an example, the particular case of an Einstein cross was considered. This reveals the entire efficiency of the application of such a CNN to detect new gravitational lenses. Depending on the properties of the observational datasets one can complement the procedures also via other software.

While currently the observations still provide not large samples of strong gravitational lenses, by time the surveys will provide more and of higher quality data and hence the 
importance for development of artificial intelligence methods for their confident detection will only increase.

\end{document}